# Base Station Network Traffic Prediction Approach Based on LMA-DeepAR

Jiachen Zhang[1*]
Key Laboratory of Trustworthy Distributed Computing and Service,
Ministry of Education
Beijing Uinversity of Posts and Telecommunications
Beijing, China
email: mailjczhang@gmail.com

Xingquan zuo[2]
Key Laboratory of Trustworthy Distributed Computing and Service,
Ministry of Education
Beijing Uinversity of Posts and Telecommunications
Beijing, China
email: zuoxq@bupt.edu.cm

Mingying xu[3]
Key Laboratory of Intelligent Telecommunication Software and Multimedia
Beijing Uinversity of Posts and Telecommunications
Beijing, China
email: xumingying0612@126.com

Jing Han[4]
Network Interface Virtualization Shanghai System Design Deptment
Zhongxing Telecommunication Equipment Corporation
Shanghai, China
email: han.jing28@zte.com.cn

Baisheng Zhang[5]
Network Interface Virtualization Shanghai System Design Deptment
Zhongxing Telecommunication Equipment Corporation
Shanghai, China
email: zhang.baisheng@zte.com.cn

*Abstract*—Accurate network traffic prediction of base station cell is very vital for the expansion and reduction of wireless devices in base station cell. The burst and uncertainty of base station cell network traffic makes the network traffic nonlinear and non-stationary, which brings challenges to the long-term prediction of network traffic. In this paper, the traffic model LMA-DeepAR for base station network is established based on DeepAR. Acordding to the distribution characteristics of network traffic, this paper proposes an artificial feature sequence calculation method based on local moving average (LMA). The feature sequence is input into DeepAR as covariant, which makes the statistical characteristics of network traffic near a period of time in the past be considered when updating parameters, and the interference of non-stationary network traffic on model training will be reduced. Experimental results show that the proposed prediction approach (LMA-DeepAR) outperforms other methods in the overall long-term prediction performance and stability of multi cell network traffic.

## I. INTRODUCTION

With the popularity of 5G network and mobile devices, people's demand for mobile data traffic is increasing. The following data traffic grows exponentially. It is very vital for network operators to provide unimpeded and ubiquitous high-quality services. Among them, how to establish an accurate long-term prediction model of base station cell network traffic to guide operators to expand the base station cell wireless equipment is a challenge.

The network traffic of base station cell is usually presented in the form of time series. The prediction period can be divided into short-term time series prediction (the prediction time is in minutes, hours and days) and long-term time series prediction (the prediction time is in days, weeks and months). The short-term traffic prediction of base station cell network is relatively mature, which mainly includes three methods, namely, traditional statistical model based methods, neural network based methods and hybrid methods combining the above two methods.

In the past, researchers have proposed to use traditional statistical models to predict network traffic such as Autoregressive Moving Average(ARMA)[1]、Autoregressive Integrated Moving Average(ARIMA)[2]. All of them are independent time series. They all set unknown parameters to fit polynomial function so as to approximate the real value of network traffic to achieve the purpose of prediction. An ARMA model based on linear least mean square error is proposed in [3] to predict network traffic; however, it requires that the network traffic must be linear and stable, and the applicability of the model is poor. Literature[4] proposed to use ARIMA model to predict network traffic, through multiple differential processing of the original sequence, and the final stationary sequence is obtained, and then ARMA is used to build the prediction model. As an improved method of ARMA, ARIMA can stabilize the original time series by differential processing, which has a wider range of applications. But it can only deal with homogeneous non-stationary time series, which still has limitations. In reference[5], SARIMA (seasonal auto regressive integrated moving average) model is proposed to predict network traffic. This model is an improved variant of ARIMA model and can fit the periodic factors of time series, but it is still poor for non-periodic and non-linear network traffic.

Neural network models for network traffic prediction include recurrent neural networks(RNN)[6], Back Propagation Neural Network(BP)[7] and its variants[8]. In reference[9], RNN is used to establish a prediction model for network traffic, which can accurately predict the traffic trend under various network application scenarios. In [10], firefly swarm algorithm is used to optimize BP neural network to predict network traffic. Simulation results show that the algorithm has high prediction accuracy. The above model has the following shortcomings in the base station cell network traffic prediction: it needs large data to learn an accurate model; it predicts for a single time series of one-dimensional variable; it does not fully consider the impact of network traffic non-stationary on the model prediction.

In recent years, hybrid model prediction method is commonly used in network traffic prediction. In order to solve the problem of self-similar network traffic prediction, a combined model of EMD and ARMA was proposed in [11]. Literature [12] is based on wavelet decomposition combined with LSTM. The performance of the combined prediction model is better than that of the ordinary LSTM network in predicting traffic burst. In [13], a method based on wavelet transform and SARIMA model is proposed, which has higher prediction accuracy than single prediction model. In [14], empirical mode decomposition (EMD) and support vector machine (SVM) are proposed to predict network traffic, which has higher accuracy than SVM alone. In [15], XGBoost combined with LSTM is proposed to predict the base station network traffic. Compared with the competitive algorithm, the proposed scheme can obtain better performance. In [16], the method of wavelet decomposition combined with BP neural network is proposed to predict the network traffic, and an adaptive learning rate method is used to optimize BP to improve the prediction accuracy. The above models are mainly used to model a single network traffic sequence and seldom consider long-term prediction of network traffic.

The base station is responsible for collecting and managing the network traffic of multiple cells. The network traffic of each cell is taken as a time series. The network traffic of the base station appears nonlinear and nonstationary, and the length of the network traffic series is often short, resulting in the long-term prediction of base station multi-cell network traffic being a challenging task.

The main contributions of this paper are as follows:

(a) Based on DeepAR model, a base station network traffic prediction approach is proposed;

(b) Aiming at the non-linear and non-stationary data of base station network traffic, the local moving average (LMA) algorithm is used to calculate the characteristic sequence to weaken the interference of non-stationary network traffic on prediction.

(c) Experimental results on real data sets show that compared with other traditional models, the LMA-DeepAR model has better prediction effect and efficiency, and the prediction accuracy is greatly improved.

The content of this paper is arranged as follows: Section 2 introduces theory and method of the proposed model; section 3 introduces the experiment and analysis; section 4 gives the conclusions.

## II. THEORETICAL METHODS

### A. DeepAR

DeepAR is a time series prediction method based on deep learning. DeepAR is composed of RNN (LSTM or GRU). The input of RNN is the sequence lag value and covariates. The training and prediction follow the general method of autoregressive model. It can effectively learn the global model from the correlated time series, and can learn complex patterns so as to predict the time series. The algorithm is described as follows[17]:

$z_{i,t}$ represents the value of time series $i$ at time $t$, given $[z_{i,1},...,z_{i,t_0-2},z_{i,t_0-1}] = \mathbf{z}_{i,1:t_0-1}$, establish the conditional probability distribution of each time sequence $[z_{i,t_0}, z_{i,t_0+1},...,z_{i,T}] = \mathbf{Z}_{i,t_0:T}$ in the future:

$$P(\mathbf{z}_{i,t_0:T} | \mathbf{z}_{i,1:t_0-1}, \mathbf{x}_{i,1:T}) \quad (1)$$

where, $t_0$ represents the time at which the prediction $z_{i,t}$ is unknown, $\mathbf{x}_{i,1:T}$ is a known covariate at all time points.

It is assumed that the model distribution $Q_\Theta(\mathbf{z}_{i,t_0:T} | \mathbf{z}_{i,1:t_0-1}, \mathbf{x}_{i,1:T})$ is composed of the product of likelihood factors:

$$Q_\Theta(\mathbf{z}_{i,t_0:T} | \mathbf{z}_{i,1:t_0-1}, \mathbf{x}_{i,1:T}) = \prod_{t=t_0}^{T} Q_\Theta(z_{i,t} | \mathbf{z}_{i,1:t-1}, \mathbf{x}_{i,1:T}) = \prod_{t=t_0}^{T} p(z_{i,t} | \theta(\mathbf{h}_{i,t}, \Theta)) \quad (2)$$

And for the parameterized $\mathbf{h}_{i,t}$ which is regarded as an autoregressive recurrent network, and is calculated by Eq. (3):

$$\mathbf{h}_{i,t} = h(\mathbf{h}_{i,t-1}, z_{i,t-1}, \mathbf{x}_{i,t}, \Theta) \quad (3)$$

where $h$ is the function of a parameterized multilayer recurrent neural network with LSTM elements. $\Theta$ is a collection of parameters. The last observation $z_{i,t-1}$ and the previous output $\mathbf{h}_{i,t-1}$ of the recurrent network will be used as the input of the next time. Likelihood function $p(z_{i,t} | \theta(\mathbf{h}_{i,t}))$ is a fixed distribution. The parameters are given by the output function $\theta(\mathbf{h}_{i,t}, \Theta)$ of the network.

Information about observation in the conditioning range of $z_{i,1:t_0-1}$ is transferred to the prediction range through the initial state $\mathbf{h}_{i,t_0-1}$. In the setting of sequence-to-sequence, the initial state is the output of the encoder network. Generally speaking, the encoder network can have different structures. Here we choose to use the same structure for the model between the condition interval and the prediction interval (corresponding to the encoder and decoder in the sequence to sequence model).

Further, we share weights between them, so that the initial state for the decoder $\mathbf{h}_{i,t_0-1}$ is obtained by computing Eq. (3) for $t = 1,...,t_0-1$, where all required quantities are observed. The initial states of both the encoder $\mathbf{h}_{i,0}$ and $z_{i,0}$ are initialized to zero.

Given the model parameters $\Theta$, we can obtain joint samples $\tilde{\mathbf{z}}_{i,t_0:T} \sim Q_\Theta(\mathbf{z}_{i,t_0:T} | \mathbf{z}_{i,1:t_0-1}, \mathbf{x}_{i,1:T})$ directly through ancestral sampling. First, we obtain $\mathbf{h}_{i,t_0-1}$ by computing Eq. (3) for $t=1,\ldots,t_0$. we sample $\tilde{z}_{i,t} \sim p(\cdot|\theta(\tilde{\mathbf{h}}_{i,t},\Theta))$ for $t=t_0,t_0+1,\ldots,T$ where $\tilde{\mathbf{h}}_{i,t} = h(\mathbf{h}_{i,t-1},\tilde{z}_{i,t-1},\mathbf{x}_{i,t},\Theta)$ initialized with $\tilde{z}_{i,t_0-1} = z_{i,t_0-1}$ and $\tilde{\mathbf{h}}_{i,t_0-1} = \mathbf{h}_{i,t_0-1}$. Samples from the model obtained in this way can then be used to compute quantities of interest, e.g. quantiles of the distribution of the sum of values for some future time period.

Figure 1 shows the training phase of DeepAR model. In each time step t, the input variables of the network are the covariates $x_{i,t}$ of the current time step and the target value $z_{i,t-1}$ and $\mathbf{h}_{i,t-1}$ of the previous step. Then, the parameters $\theta_{i,t}$ of the likelihood function $p(z|\theta)$ are calculated by Eq. (4).

$$\theta_{i,t} = \theta(\mathbf{h}_{i,t},\Theta) \quad (4)$$

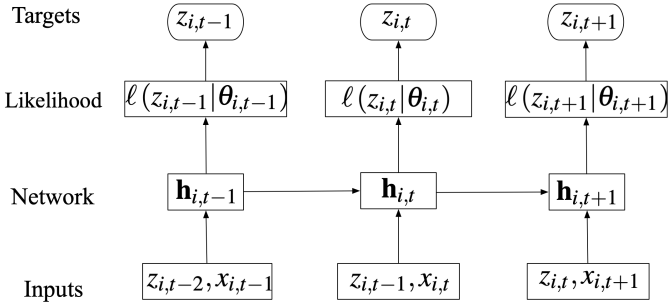

Fig 1. Training process of DeepAR

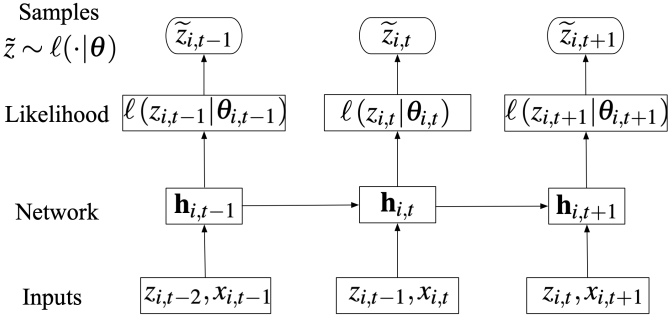

Fig 2. Prediction process of DeepAR

Figure 2 shows the model prediction stage. Import historical data of $t<t_0$ from time series $z_{i,t}$. Then randomly collect samples at $t \geq t_0$ to obtain $\hat{z}_{i,t} \sim p(\cdot|\theta_{i,t})$ and feedback to the next time step until the sample trajectory is generated at the end of the prediction range. Repeat this step to generate many trajectories representing the joint prediction distribution. Improved prediction method.

Due to the influence of various external factors on the user behavior of network traffic, the network behavior has certain regularity and contingency, and the nonlinear macro network traffic has certain regularity, suddenness and contingency. According to these characteristics of traffic behavior, network traffic $X(t)$ can be decomposed into deterministic components $y(t)$, which include trend part, periodic part, burst part and random part $e(t)$[18].

$$X(t) = y(t) + e(t) \quad (5)$$

In order to restrain the influence of the random part $e(t)$ of time series on $X(t)$, the data is smoothed, and the non-stationary series $X(t)$ can be regarded as nearly stationary in the small time step interval from $t_i$ to $t_{i+n}$.

TABLE I.  LOCAL MOVING AVERAGE ALORITHM

| Algorithm 1 Local Moving Average Algorithm |
|---|
| **Input**: Time series training data $\mathbf{Z} = \{z_1, z_2, \ldots, z_n\}$, training input length *cl*, prediction length *pl*, and *cl* not less than *pl*, maxminum characteristic class num *K*, training data series length *n*. |
| **Output**: Characteristic series $\mathbf{X}_k = \{\mathbf{x}_1, \mathbf{x}_2, \ldots, \mathbf{x}_K\}$. |
| **Begin:** |
|   1: set *v*:=0. |
|   2: **for** *k* = 1 to *K* **do** |
|   3:   **for** *i* = 1 to *n* + *pl* **do** |
|   4:     **if** *i* < *pl* **then** |
|         Compute the series $z[1,2,\ldots,cl]$ feature value *v*, set $x_i := v$, update $v := 0$. |
|   5:     **else if** $pl \leq i < n$ **then** |
|   6:       **if** *n-i*-1 < *pl* **then** |
|         Compute the seires $z[(n-cl),\ldots,n]$ feature value *v*, set $x_i := v$, update *v*:=0. |
|   7:       **else** |
|         Compute the series $z[(i-pl+1),\ldots,(cl+i-pl+1)]$ feature value *v*, set $x_i := v$, update *v*:=0. |
|   8:       **end if** |
|   9:     **else** |
|         Compute the series $z[n-pl,\ldots,n]$ feature value *v*, and set $x_i := v$, update *v*:=0. |
| 10:     **end if** |
| 11:   **end for** |
|       set $\mathbf{X}_k := \mathbf{x}$. |
| 12: **end for** |

Take the average value of each *n* adjacent data to represent the value of any one of the *n* data, and regard it as the measurement result that suppresses the random error or the signal that eliminates the noise[19]. In DeepAR model, both the covariate $\mathbf{x}_i$ and the target variable $\mathbf{z}_i$ are the input variables of the model, and the covariate $\mathbf{x}_i$ depends on $\mathbf{z}_i$ and changes with time. For example, in the prediction of commodity sales volume, the covariate can be used as the variable series that affects the sales volume[17].

Inspired by the above, in the base station network traffic prediction, this paper takes the covariate as an artificial characteristic variable, such as the traffic distribution characteristics of a certain time interval. Therefore, this paper introduces a local moving average (LMA) feature sequence calculation method.

The feature sequences calculated by this method are input into the model as the covariates of the DeepAR model to improve the training accuracy of the model. Because different cells are distributed in different physical spaces, the distribution characteristics of network traffic collected by different cells are not the same. Therefore, for accurate prediction of each cell, it is necessary to calculate the characteristic sequence through the local moving average algorithm. The algorithm description is shown in table 1. Among them, for the calculation of $v$ in the algorithm, this paper calculates the average and standard deviation of the sequence $z_i$.

## III. EXPERIMENT AND ANALYSIS

### A. Model evaluation measure

Because of the large traffic order of base station network, this paper uses RMSLE as the metric, a smaller value of the metric indicates the better the prediction effect. RMSLE is calculated by Eq. (6).

$$RMSLE = \sqrt{\frac{1}{N}\sum_{i=1}^{N}\left(\log(\hat{f}+1)-\log(f_i+1)\right)^2} \quad (6)$$

where, $\hat{f}$ is the predicted value of the model and $f$ is the real value. N is the product of the day step size of the traffic to be predicted and the number of base station network traffic series.

### B. Experimental environment

Hardware: GTX 1070 Ti Graphics 8G E5-2678 v3 24 cores Memory 64G 220GB SSD + 6TB HDD.

Software environment: tensorflow(2.3) keras(2.0.3) python(3.7) xgboost(1.2.1) mxnet(1.6.0), gluonts(0.5.0), R(3.6.1), statsmodels(0.11.1), sklearn(0.23.1).

Experimental setup: in this paper, LMA-DeepAR is compared with six models, ETS, ARIMA, XGBOOST, LSTM, and DeepAR. The ETS and ARIMA models are implemented using the forecast package in R[20] where the ARIMA model uses the auto.arima() function to automatically select the model best parameters. When using the ETS model, the additive only parameter is set to be TRUE. XGBoost model is implemented using the xgboost[21] package, with the parameter eval_metric: the evaluation criterion during model fitting, which is set to rmse, and the parameter n_estimators: set to be 1000 as the number of learners parameter. LSTM model is implemented using TensorFlow[22] and Keras[23], the input step parameter is set to be 62, a three-layer neural network architecture is used, and the learning rate is set to be 0.01. The deep autoregressive model (DeepAR) is implemented using GluonTS[24]. The hyperparameters involved in DeepAR are context_length; the number of context steps, which is used as the input step of the model and is set to be 62; epoch which is the number of times to train the complete data, which is set to be 15; dynamic_feat-the artificial features (including standard deviation, mean, etc.), which are calculated using the LMA algorithm and are used as model covariates input sequence.

TABLE II. RMSLE WITH DIFFERENT STEP SIZES FOR DIFFERENT MODELS

| Time-Step | ETS | ARIMA | XGBoost | LSTM | DeepAR | LMA-DeepAR |
|---|---|---|---|---|---|---|
| 15 | 1.4105 | 1.4221 | 1.3431 | 1.4721 | 1.2966 | 1.1543 |
| 16 | 1.4146 | 1.4232 | 1.3459 | 1.4713 | 1.2997 | 1.1477 |
| 17 | 1.4235 | 1.4296 | 1.3514 | 1.4795 | 1.3065 | 1.1463 |
| 18 | 1.4263 | 1.4306 | 1.3535 | 1.4804 | 1.3152 | 1.1515 |
| 19 | 1.4257 | 1.4315 | 1.3522 | 1.4786 | 1.3184 | 1.1475 |
| 20 | 1.4307 | 1.4356 | 1.3515 | 1.4774 | 1.3249 | 1.1486 |
| 21 | 1.4334 | 1.4412 | 1.3513 | 1.4795 | 1.3321 | 1.1513 |
| 22 | 1.4399 | 1.4483 | 1.3561 | 1.4843 | 1.3426 | 1.1531 |
| 23 | 1.4434 | 1.4519 | 1.3581 | 1.4851 | 1.3509 | 1.1556 |
| 24 | 1.4542 | 1.4606 | 1.3671 | 1.495 | 1.3617 | 1.1602 |
| 25 | 1.4647 | 1.4683 | 1.3723 | 1.504 | 1.372 | 1.1666 |
| 26 | 1.4678 | 1.4717 | 1.3736 | 1.5065 | 1.38 | 1.1673 |
| 27 | 1.4726 | 1.4759 | 1.3757 | 1.5097 | 1.3882 | 1.1689 |
| 28 | 1.4798 | 1.4812 | 1.3787 | 1.5139 | 1.3973 | 1.1736 |
| 29 | 1.4844 | 1.4858 | 1.3842 | 1.5176 | 1.4074 | 1.1806 |
| 30 | 1.4931 | 1.4954 | 1.3951 | 1.5254 | 1.4195 | 1.1885 |
| 31 | 1.5031 | 1.5051 | 1.4075 | 1.5356 | 1.4316 | 1.198 |
| mean | 1.451 | 1.4564 | 1.3657 | 1.4951 | 1.3556 | 1.1623 |

### C. Analysis of experimental results

The network traffic collected from a base station in a city in China which is used as the object of this experiment, and there are 1000 cell network traffic under this base station. The number unit and the collection duration of the network traffic of each cell are the same. The unit of traffic data is KB/day (by day), and the daily traffic data is the current self-busy traffic value, the so-called self-busy time refers to the maximum traffic hour of the day. The traffic data is collected from September 1, 2017 to March 31, 2018, and the traffic data in the time range of

2017/9/1-2018/2/28 is selected as the training set of the model, and the traffic data in the range of 2018/3/1-2018/3/31 is selected as the test set of the model. For the base station wireless device expansion and contraction scenario, the prediction step 15-31 interval is specified in this paper long-term prediction. In order to better compare and analyse the performance differences between different models, two experiments are conducted.

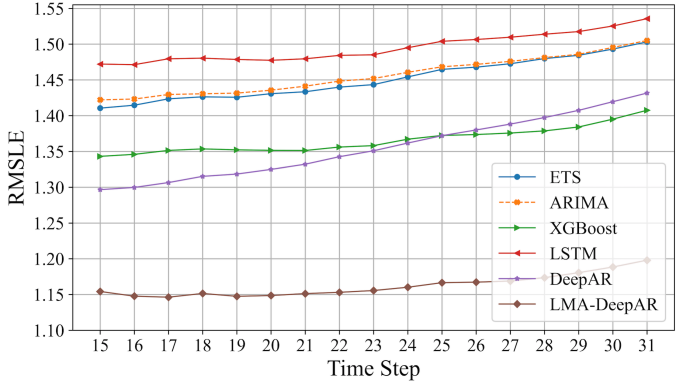

Fig 3. Comparison of RMSLE trends predicted by different models with different step sizes

Experimental analysis I: The network traffic of 1000 base station cells is predicted using different models with prediction steps of 15-31, and then all the predicted results of 1000 cells at each step are taken as a whole and the corresponding RMSLE results at each step are calculated using Eq. (6). Figure 3 represents the prediction results of different models at each step in long-term prediction. The accuracy of each model in long-term prediction can be obtained by averaging the prediction results of all steps, and we can see from table 2 that the LMA-DeepAR model has the best average prediction performance, and the DeepAR model has the second best average prediction performance, and comparing the average prediction results of the LMA-DeepAR model and the DeepAR model, we can see that the introduction of the LMA method can greatly improve the The prediction accuracy of DeepAR model in the long-term prediction of network traffic of base station cells.

Figure 3 shows the trend changes of the predicted RMSLE values. In the domain of long-term prediction of base station cell network traffic, in the interval of step 15 to 22, with the increase of step size, the RMSLE values corresponding to different models except LMA-DeepAR model increase slowly, while when the prediction step size is larger than 22, the RMSLE of all models increase gradually and the increase becomes larger, which reflects that the long-term prediction of cell traffic at the base station is a rather challenging task.

Experimental analysis II: In experimental analysis I, for each step, computing an RMSLE value by taking the prediction results of network traffic of all 1000 cells as a whole, we are able to analyze the overall prediction performance of all cells under different steps for different models. However, when some of the cells are predicted very poorly, if they are taken as part of the overall prediction results of all cells, it will seriously affect the overall prediction results, and indirectly indicates that the model lacks stability in prediction performance. The so-called prediction performance stability refers to the size of fluctuation between the prediction results of each of the 1000 cells and the average of the prediction results of all cells, and we measure this index by calculating the standard deviation of the RMSLE values of the 1000 cells. To illustrate the stability of the prediction performance of different models and to show that the model prediction effect is more statistically significant, further analysis of the experimental results is needed. The model prediction performance stability analysis process is as follows:

In this experiment, there are a total of 1000 base station cell network traffic. Each model in different prediction steps need predicting the traffic of each cell and getting the prediction results. With the experimental analysis of all cells under each step to calculate a RMSLE value, a RMSLE value is calculated in each step of each cell. This value only reflects the prediction performance of the model in a single cell and cannot be used as a reference indicator to guide the expansion and contraction of the base station. But it can be used to analyze whether the prediction performance of the model is stable.

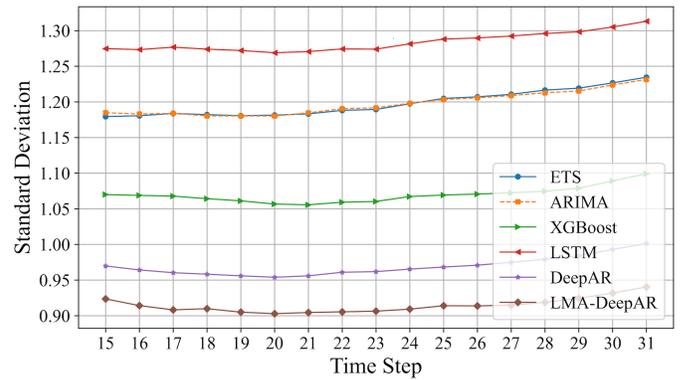

Fig 4. Comparison of standard deviation results for different models with different step sizes

For each model to obtain 1000 RMSLE results at each step, in order to analyze the stability of the model prediction performance: (1) First, the standard deviation of the mean of all 1000 cell RMSLEs for each model at each step is calculated (as shown in Figure 4). (2) The standard deviations of the mean values of RMSLE for all 1000 cells of the model at each step were calculated for long-term prediction (step 15 to step 31) at their corresponding steps, respectively. The statistical results are shown in table 3, As shown in Table 3, LMA-DeepAR still performs best in terms of stability of model prediction performance.

TABLE III. PREDICTION OF STANDARD DEVIATION UNDER DIFFERENT MODELS WITH DIFFERENT STEP SIZES

| Time-Step | ETS | ARIMA | XGBoost | LSTM | DeepAR | LMA-DeepAR |
|---|---|---|---|---|---|---|
| 15 | 1.17931 | 1.18481 | 1.06992 | 1.27473 | 0.96976 | 0.92353 |
| 16 | 1.18053 | 1.18293 | 1.06875 | 1.27331 | 0.96416 | 0.91415 |
| 17 | 1.18375 | 1.18381 | 1.06776 | 1.27673 | 0.96028 | 0.90814 |
| 18 | 1.18211 | 1.1803 | 1.06412 | 1.27395 | 0.95817 | 0.90982 |
| 19 | 1.18055 | 1.18032 | 1.06109 | 1.27205 | 0.9559 | 0.90496 |
| 20 | 1.18142 | 1.18011 | 1.05663 | 1.26891 | 0.95387 | 0.90282 |
| 21 | 1.18314 | 1.18459 | 1.05553 | 1.27056 | 0.9559 | 0.90442 |
| 22 | 1.18807 | 1.19044 | 1.05919 | 1.27425 | 0.96092 | 0.90538 |
| 23 | 1.18937 | 1.19166 | 1.06025 | 1.27397 | 0.96179 | 0.90633 |
| 24 | 1.19729 | 1.1979 | 1.06718 | 1.28143 | 0.96532 | 0.90917 |
| 25 | 1.20484 | 1.20333 | 1.06928 | 1.28802 | 0.96822 | 0.91395 |
| 26 | 1.207 | 1.20568 | 1.07059 | 1.28968 | 0.97082 | 0.91372 |
| 27 | 1.21057 | 1.20879 | 1.0724 | 1.29233 | 0.9748 | 0.91468 |
| 28 | 1.2165 | 1.21275 | 1.07464 | 1.29595 | 0.97919 | 0.91872 |
| 29 | 1.21918 | 1.21523 | 1.07888 | 1.29826 | 0.98412 | 0.92434 |
| 30 | 1.22661 | 1.22368 | 1.08905 | 1.30498 | 0.99298 | 0.93176 |
| 31 | 1.23449 | 1.2314 | 1.09916 | 1.31307 | 1.00119 | 0.94028 |
| mean | 1.1979 | 1.1975 | 1.0697 | 1.2837 | 0.9693 | 0.9145 |

Figure 4 shows that LMA-DeepAR has the best prediction stability in the long-term prediction, and the DeepAR model shows the second best prediction stability. In Figure 3 when the prediction step size is greater than 25 the LSTM model will outperform DeepAR, although this does not mean that the LSTM prediction performance is good. We can see from Figure 4 that the stability of the prediction performance of the DeepAR model is significantly better than the other models in the prediction of cell network traffic at the base station, except for the LMA-DeepAR model.

ACKNOWLEDGMENTS

This work was supported by the ZTE Industry University Research Cooperation Project(NO. 2018zte01-02-09).

IV. CONCLUSION

Both the prediction accuracy and stability of the model are equally important in the long-term prediction of network traffic in base station cells. In this paper, we consider the special characteristics of base station cell network traffic prediction and use artificial features as covariates of the model for better long-term prediction based on DeepAR model, and propose a local sliding average calculation method to calculate the artificial feature variables based on the cell traffic distribution characteristics, through getting the feature sequences as the covariate sequences of the model to input into the model. The comparison experiments show that LMA-DeepAR is effective in predicting the base station cell network traffic. Compared with ETS, ARIMA, XGBoost, LSTM and DeepAR, and LMA-DeepAR shows the best accuracy and prediction stability in predicting base station cell network traffic.